\documentclass{optica-article}

\journal{opticajournal} 

\articletype{Research Article}

\usepackage{lineno}
\usepackage{bm}
\usepackage{siunitx}

\def\Vecu{\mathbf{u}}

\def\VecT{\mathbf{T}}

\begin{document}

\title{Fast and light-efficient wavefront shaping with a MEMS phase-only light modulator}

\author{
José~C.~A.~Rocha\authormark{1,2,*}, Terry Wright\authormark{3}, Un\.e~G.~B\=utait\.e\authormark{1}, Joel~Carpenter\authormark{2}, George~S.~D.~Gordon\authormark{3} \& David~B.~Phillips\authormark{1,$\dagger$}}

\email{\authormark{*}jd964@exeter.ac.uk} \email{\authormark{$\dagger$}d.phillips@exeter.ac.uk}


\address{
\authormark{1}Department of Physics and Astronomy, University of Exeter, Exeter, EX4 4QL. United Kingdom.\\
\authormark{2}School of Electrical Engineering and Computer Science, The University of Queensland, Brisbane, QLD 4072, Australia.\\
\authormark{3}Optics and Photonics Group, University of Nottingham, Nottingham, NG7 2RD, United Kingdom.
}

\begin{abstract*}
Over the last two decades, spatial light modulators (SLMs) have revolutionised our ability to shape optical fields. They grant independent dynamic control over thousands of degrees-of-freedom within a single light beam. In this work we test a new type of SLM, known as a {\it phase-only light modulator} (PLM), that blends the high efficiency of liquid crystal SLMs with the fast switching rates of binary digital micro-mirror devices (DMDs). A PLM has a 2D mega-pixel array of micro-mirrors. The vertical height of each micro-mirror can be independently adjusted with 4-bit precision. Here we provide a concise tutorial on the operation and calibration of a PLM.
We demonstrate arbitrary pattern projection, aberration correction, and control of light transport through complex media. We show high-speed wavefront shaping through a multimode optical fibre -- scanning over 2000 points at 1.44\,kHz. We make available our custom high-speed PLM control software library developed in C++. As PLMs are based upon micro-electromechanical system (MEMS) technology, they are polarisation agnostic, and possess fundamental switching rate limitations equivalent to that of DMDs -- with operation at up to 10\,kHz anticipated in the near future. We expect PLMs will find high-speed light shaping applications across a range of fields including adaptive optics, microscopy, optogenetics and quantum optics.
\end{abstract*}


\section{Introduction}
Spatial light modulators (SLMs) are planar optical devices capable of dynamically shaping light fields. They have a broad array of applications, ranging from holography and adaptive optics~\cite{vellekoop2007focusing,hampson2021adaptive}, to microscopy~\cite{maurer2011spatial}, optical tweezers~\cite{grier2003revolution}, optogenetics~\cite{emiliani2022optogenetics}, wavelength selective switching~\cite{ma2020recent},
and emerging 3D augmented reality displays~\cite{gopakumar2024full}. In this work we test a newly developed type of fast-switching SLM, known as a {\it phase-only light modulator} (PLM)~\cite{bartlett2021recent,douglass2022reliability,byrum2024optimizing}, and assess its performance at a variety of challenging light shaping tasks. To put this new device into context, we first summarise the capabilities of well-established SLM technology and outline how PLMs are poised to fill a critical performance gap.

At present, the most widely used commercially available SLMs fall into three main categories: 
liquid crystal-based devices, deformable mirrors and digital micromirror devices (DMDs). 
Liquid crystal SLMs impart a tunable phase delay to one linear polarisation state of the reflected (or transmitted) light. 
An incident optical field passes through a liquid crystal layer sandwiched between a pair of electrodes. The refractive index of this layer is varied by adjusting the voltage applied across it, which in turn shifts the optical path length of the cell~\cite{lazarev2019beyond}. Liquid crystal SLMs offer high efficiency (>80\,$\%$) and high bit depth phase modulation (8-bit to 16-bit) with resolutions up to $10$ million pixels, and a pitch down to $\sim$\SI{4}{\micro\meter}~\cite{yang2023review}. 
A key limitation of liquid crystal SLMs is their relatively low switching rates, due to the slow response time of the viscous liquid crystal layer. Standard models typically switch at video rates (60-120\,Hz). More recently, high-speed liquid crystal SLMs have been developed with switching rates of $\sim0.4-1.7$\,kHz, typically calculated using the time taken to switch from 10\% to 90\% of the desired phase change (so the fully settled switching rates are roughly two thirds of these rates), and the very highest rates are possible only for shorter wavelengths ($<$550\,nm). These enhanced switching rates are attained by heating the liquid crystal to reduce its viscosity, or applying overdrive schemes to exert higher instantaneous torques on the liquid crystal molecules~\cite{thalhammer2013speeding}. Ferroelectric liquid crystal SLMs can have switching rates in excess of 3\,kHz, but at the cost of binary phase modulation resulting in low diffraction efficiency ($<$50\%), due to the Hermitian symmetry of the Fourier transform of a real function~\cite{hossack2003high,maurer2008suppression}.

A second category of SLMs is the deformable mirror, consisting of an array of micro-actuators that locally change the height of a reflective mirrored surface. 
State-of-the-art deformable mirrors have up to a few thousand independent actuators, each capable of vertically pistoning by a distance of several wavelengths at switching rates up to a few kilohertz~\cite{bifano2011mems,madec2012overview}.
However, such high-end deformable mirrors are typically bespoke devices which are prohibitively expensive, restricting their wider application. Lower specification deformable mirrors, with tens to a few hundred actuators, and switching speeds of hundreds of Hertz, find applications correcting for weak aberrations in microscopy~\cite{booth2002adaptive,bourgenot20123d}. In general, deformable mirrors are well-suited for correcting weak phase aberrations, but lack the resolution necessary for intricate beam shaping.

A third type of SLM is the DMD -- rapidly reconfigurable binary intensity modulators consisting of mega-pixel arrays of micro-mirrors. Each mirrored pixel can be tilted between two states at switching rates in excess of 20\,kHz.
DMDs can encode complex holograms to shape both the phase and intensity of an incident beam projected into the far-field
\cite{brown1966complex,conkey2012high,popoff2023practical}. However, the binary amplitude nature of DMDs, and the large number of diffraction orders generated by diffraction from the edges of the tilted pixels, renders this form of beam shaping very inefficient, with typically <1\% of light being usefully controlled~\cite{turtaev2017comparison}, and lowers the effective number of controllable degrees-of-freedom by several orders of magnitude~\cite{goorden2014superpixel,gutierrez2023binary}.\\

\noindent {\it The Phase-only Light Modulator:} As outlined above, well-established SLM technology struggles to simultaneously deliver efficient mega-pixel resolution light shaping at rapid switching rates. This gap has motivated the exploration of a variety of new beam shaping platforms, which have exciting promise for the future~\cite{sun2013large,tzang2019wavefront,li2019phase,feldkhun2019focusing,ersumo2020micromirror,mounaix2020time,panuski2022full}.  
Recently, {\it Texas Instruments} (TI) have developed the PLM: a new addition to the SLM family that offers high-speed light shaping capabilities beyond that of the pre-existing technology~\cite{bartlett2021recent,douglass2022reliability,byrum2024optimizing}.

\begin{figure*}[t]
\centering\includegraphics[width=\linewidth]{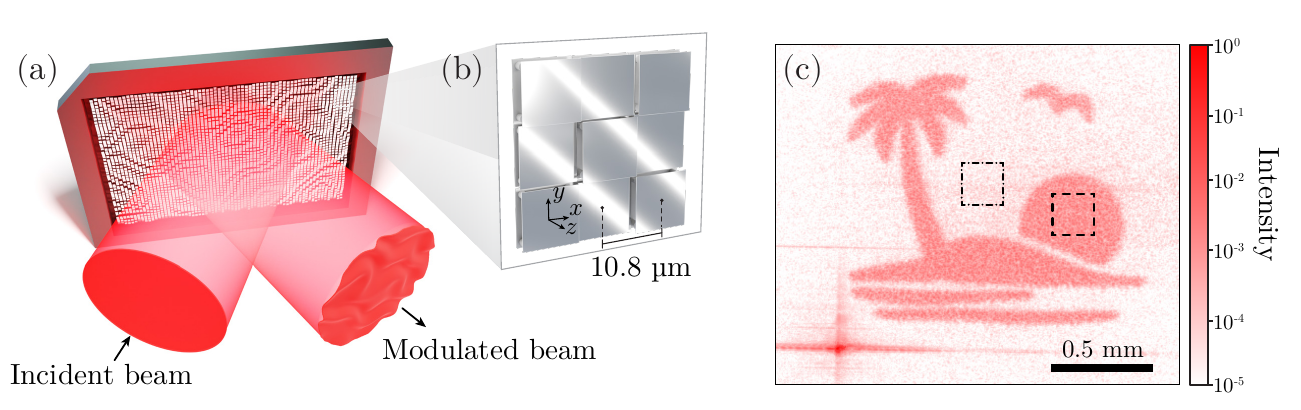}
\caption{{\bf The phase-only light modulator}. (a) A schematic of a PLM. A collimated incident laser beam diffracts from the structured PLM surface. (b) A close-up view of a $3\times3$ array of micro-mirrors. 
(c) Pattern projection using a PLM. Experimentally measured high-dynamic range composite image, represented on a log scale, of a palm tree scene projected into the Fourier plane of the device (using a lens of focal length 300\,mm and subsequent demagnification factor of 3$\times$). The PLM phase hologram is designed using the Gerchberg-Saxton algorithm~\cite{yang1994gerchberg}. The image contains $\sim$70\% of the diffracted light. The contrast is $\sim$37, given by the ratio of average intensity of the bright and dark parts of the scene, measured within the two outlined boxes. Unmodulated zero-diffraction order light is visible in the bottom-left corner.}
\label{fig:mainfig}
\end{figure*}

PLMs are micro-electromechanical system (MEMS)-based SLMs consisting of mega-pixel arrays of micro-mirrors, as shown schematically in Fig.~\ref{fig:mainfig}. Each micro-mirror is mounted on an electrostatically driven piston that can independently adjust the mirror's vertical displacement with 4-bit precision (i.e.~16 mirror heights). Micro-mirror response time is less than \SI{50}{\micro\second}, yielding a fundamental switching rate of $\sim$20\,kHz -- although the currently available test models are limited to 1.44\,kHz by their control electronics. PLMs are based on the same underlying TI Digital Light Processing (DLP) technology as DMDs (also developed by TI)
enabling the adapation of pre-existing driving hardware and software. An arrangement of 4 individually addressable electrodes sits behind each micro-mirror. Each unique combination of activated electrodes moves the height of the micro-mirror to a different vertical position~\cite{bartlett2019adapting,oden2020innovations}. Current PLM models are designed to operate at visible wavelengths, and can impart phase shifts of up to $2\pi$\,rad to light of wavelengths $\lambda$ over the range $405$\,nm$ <\lambda< 650$\,nm (with a maximum change in micro-mirror height of 325\,nm). Available PLMs have $\sim$$10^6$ micro-mirrors, with a pixel pitch of $\sim$\SI{11}{\micro\meter} and an area fill factor of $\sim$95\%, resulting in a high diffraction efficiency rivalling that of liquid crystal SLMs~\cite{bartlett2021recent,douglass2022reliability,byrum2024optimizing}.

Here we provide a concise tutorial on the operation and calibration of a PLM, test its performance at a range of light shaping tasks, and make available our custom high-speed PLM control software. We demonstrate how to optimally tune the phase shift range for a given illumination wavelength, calibrate the 4-bit phase response for use in both the near and far-field of the device, and synchronise the data-upload to display holograms continuously at 1.44\,kHz. We measure and correct for the curvature of the PLM chip itself, and show arbitrary pattern projection, simultaneous modulation of both amplitude and phase of light, and precise high-speed spatial control of light transport through a complex multiple scattering medium -- a multimode optical fibre. Our work demonstrates that PLMs can deliver high efficiency and high-fidelity wavefront shaping on-par with state-of-the-art SLMs, in a rapid switching platform.

\section{PLM specification and setup}

\begin{figure*}[h]
\centering\includegraphics[width=\linewidth]{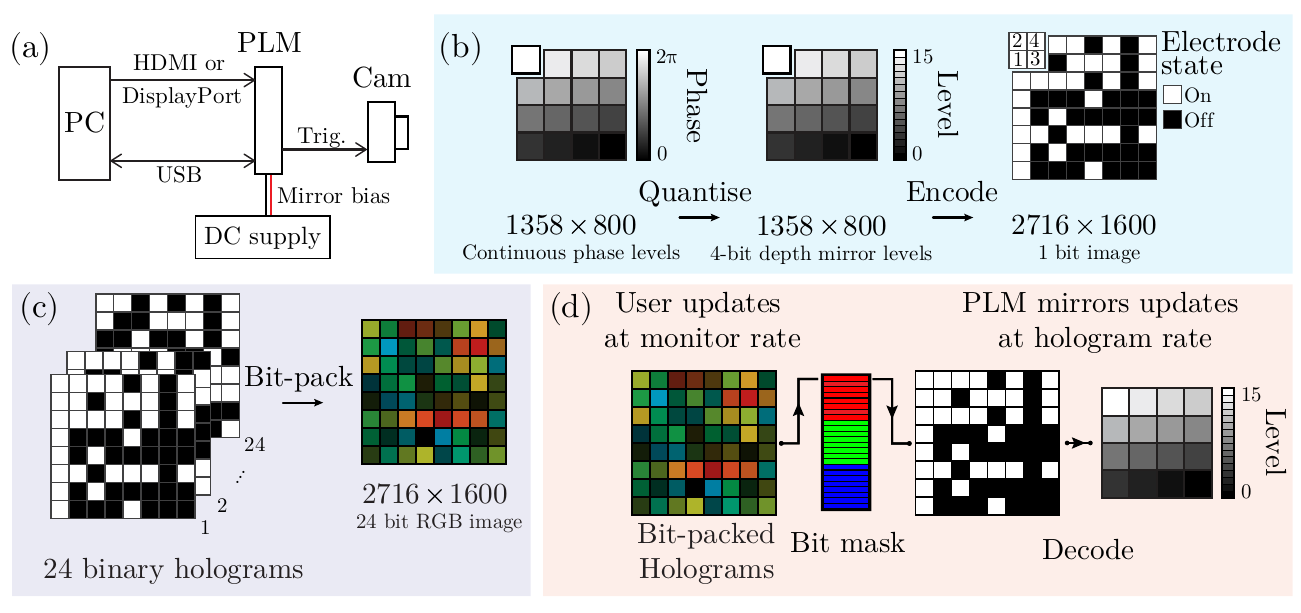}
\caption{{\bf PLM setup and hologram display process}. (a) Schematic of PLM connections. (b) {\it Single hologram creation}: A continuous phase level hologram is quantised to the available 4-bit depth mirror levels. A $4\times 4$ micro-mirror section is shown for clarity. Each 4-bit mirror level is spatially encoded into a $2\times 2$ binary element `memory cell' representing the on/off states of the 4 electrodes (labeled from 1 to 4) behind each micro-mirror. (c) {\it Bit packing}: 24 binary holograms are combined into one 24-bit RGB image by placing each hologram into the bits of the 3 colour channels (8 bits per color channel). (d) {\it Display and update}: The user updates bit-packed holograms at the monitor rate (30 or 60\,Hz). The PLM extracts individual holograms using a bit mask, decodes the binary pattern to set mirror array heights, and updates mirrors at the hologram rate ($24\times$ the monitor rate).
}
\label{fig:holo}
\end{figure*}

We test an evaluation model PLM (DLP6750 EVM) featuring a resolution of $1358\times800$ micro-mirrors. The PLM is controlled via a USB connection, and data designating the desired micro-mirror states is streamed via a High Definition Multimedia Interface (HDMI) or DisplayPort interface. In both cases, upon connection the computer recognises the PLM as an additional monitor. Figure~\ref{fig:holo}(a) shows a schematic detailing the PLM connections. The state of each PLM mirror is encoded within a $2\times 2$ block of pixels -- hence the computer interprets the PLM as a monitor of dimension $2716\times1600$ pixels. Each single frame transmitted across the HDMI or DisplayPort connection encodes the data for the PLM to sequentially display 24 phase holograms.

Figure~\ref{fig:holo}(b) illustrates the procedure to generate a single PLM hologram. Continuous phase values are quantised according to the phase-shift imparted by each mirror-level -- which is first calibrated and stored in a lookup table (see calibration methods described below). An individual PLM hologram is encoded in a binary $2716\times1600$ pixel array, where each group of $2\times2$ pixels -- known as a {\it memory cell} -- spatially encodes a 4-bit number defining the level of the corresponding micro-mirror. As shown in Fig.~\ref{fig:holo}(c), 24 of these binary images are packed together into a single full-colour red-green-blue (RGB) frame having 8-bits per colour channel. Upon receiving an RGB frame, the PLM will unpack this data by bit masking the color channels, and cycle through the 24 holograms (Fig.~\ref{fig:holo}(d)).

In our evaluation model PLM, each hologram is displayed for \SI{554}{\micro\second}, with a delay of \SI{140}{\micro\second} between adjacent holograms in the set. This delay is divided into a \SI{90}{\micro\second} time window in which the micro-mirrors return to a flat configuration while the next hologram is loaded to the chip, followed by \SI{50}{\micro\second} during which the micro-mirrors move to their new positions~\cite{gong2024exploring}. We note that later releases of PLM firmware are expected to reduce or completely remove the \SI{90}{\micro\second} return to flat between frames. Using the 60\,Hz DisplayPort interface, the PLM is capable of displaying holograms at a rate of $24\times60$\,Hz = 1.44\,kHz. If using the 30\,Hz HDMI connection, the hologram display rate is $24\times30$\,Hz = 720\,Hz. In this case each hologram is displayed twice in succession.

Output trigger lines are provided that transmit digital pulses at various points in the hologram display cycle, enabling synchronisation with external equipment. 
In this work we utilise an output trigger line that emits a pulse each time a new hologram is displayed to synchronise the image capture of a high-speed camera with the PLM at up to 1.44\,kHz. We note that, in principle, it is possible to display a small number of preloaded holograms at a rate of 5.76\,kHz using this evaluation model, although this operation is not yet fully supported and so we were not able to test this functionality. Furthermore, 10\,kHz operation is planned for the near future, relying on a custom controller~\cite{oden2020innovations,bartlett2021recent}.

PLM initialisation is achieved using a graphical user interface (LightCrafterDLPC900). This application initialises the device, allows the bit-unpacking order and endianness to be configured, and the hologram display time to be adjusted (with a minimum time of \SI{554}{\micro\second} at present). We have developed libraries of control functions in C++ allowing PLM operation to be automated and data streaming to be synchronised for continuous operation. These libraries can be called from top-level PLM control software in another language -- we have tested Python, MATLAB, and LabVIEW.

\section{PLM calibration}
The range of heights over which the micro-mirrors must move to impart a $2\pi$ phase shift is wavelength dependent and given by $\lambda/2$. Therefore, PLMs are equipped with a control line input (known as the `mirror bias') that enables the mirror height range to be tuned as a function of the supplied voltage (see Fig.~\ref{fig:holo}(a)). This allows the PLM response to be optimised to the illumination wavelength.

It is also necessary to calibrate the phase delay imparted to reflected light as a function of the mirror state. Labelling the mirror states with index $i$, numbered from 1 to 16, the vertical displacement $d(i)$ of the micro-mirrors is a non-linear function of $i$. The phase delay imparted to reflected light, $\theta(i)$, is also wavelength dependent: $\theta(i) = 4\pi d(i)/\lambda$. PLM calibration is achieved by measuring this phase delay as a function of micro-mirror state, which is stored in a lookup table for a given illumination wavelength. Continuous values of the required phase delay at each pixel in the desired hologram are then rounded to the nearest achievable phase delay according to this lookup table, setting the associated micro-mirror state.

\subsection{Image plane calibration}
We first demonstrate the determination of the optimal mirror bias and lookup table, at a given wavelength, by directly imaging the plane of the PLM chip. We set up a Twyman-Green interferometer as shown in Fig.~\ref{fig:calib}(a). A linearly polarised collimated laser beam of wavelength ${\lambda=633\,\text{nm}}$ is split into two paths of equal length by a non-polarising beam-splitter. One path reflects from the PLM chip, and the second reference path reflects from a mirror (of flatness $\sim$$\lambda/10$ in this case). The reflected light is recombined at the same beam-splitter and directed onto a camera (Cam 2, Ximea xiQ  M0013MG-E2) imaging the surface of the PLM. This arrangement produces an interferogram where the fringes display a `contour map' of the surface height of the PLM -- providing a convenient way to directly visualise changes in mirror heights.

By tilting the reference arm mirror a small amount, the reference beam arrives at the camera at a small angle with respect to the light reflecting from the PLM, resulting in the fringe pattern on the camera becoming dominated by stripes. To directly visualise how the phase of reflected light changes with micro-mirror height, we piston the micro-mirrors through their 16 states, causing the interference fringes to translate sideways. The imparted phase shift as a function of micro-mirror state, $\theta(i)$, can then be measured by fitting a sinusoid to the interference stripes and tracking the change in phase of this sinusoid as it moves sideways. For example, the inset in Fig.~\ref{fig:calib}(a) shows the interference stripes when the micro-mirrors on the right-hand-side of the PLM chip are set to change the phase of reflected light by $\pi$\,rad with respect to the light reflected from the left-hand-side of the PLM chip. During lookup table calibration, we found that the states may need reordering so that the phase monotonically increases. Repeating this lookup table calibration for different mirror bias voltages enables the optimum mirror bias to be identified. Figure \ref{fig:calib}(b) shows a plot of the achievable phase range (maximum phase shift minus minimum phase shift) as a function of mirror bias voltage. Figure~\ref{fig:calib}(c) shows an example of the measured lookup table with an optimal mirror bias (of -0.5\,V for ${\lambda=633\,\text{nm}}$) yielding $2\pi$ phase control of reflected light.

\subsection{Fourier plane calibration} 
\begin{figure*}[b]
\centering\includegraphics[width=\linewidth]{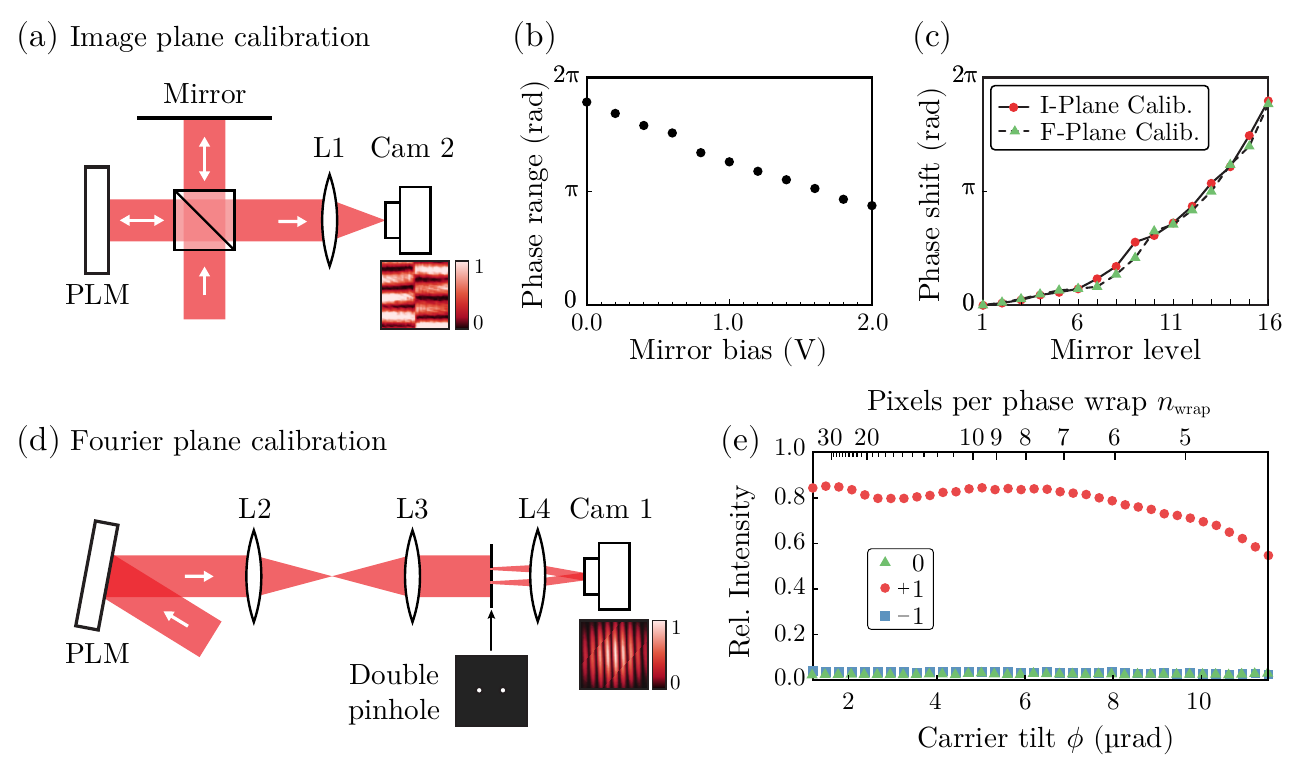}
\caption{{\bf Calibration and diffraction efficiency}. (a) A schematic of the Twyman-Green interferometer for image plane calibration. $\text{L1}=150$\,mm focal length. (b) Achievable phase shift range as a function of mirror bias voltage.  (c) PLM lookup tables measured using both the image plane and Fourier plane setups, showing very similar calibration curves. (d) Experimental setup for Fourier plane PLM calibration. Focal lengths: $\text{L2}=300$\,mm; $\text{L3}=200$\,mm; $\text{L4}=175$\,mm. (e) The measured diffraction efficiency of the calibrated PLM as a function of carrier tilt $\phi$. The setup for this experiment is the same as (d), but with the double pinhole removed. The relative intensity of light sent to the +1, 0, and -1 diffraction orders is plotted. 
}
\label{fig:calib}
\end{figure*}

SLMs are often used diffractively to shape light in the far-field of the modulator chip. It is therefore useful to be able to characterise the lookup table and tune the mirror bias voltage by observing the Fourier plane of the PLM. Figure~\ref{fig:calib}(d) shows a setup to achieve this. We illuminate the PLM with a 1\,mW linearly polarised collimated beam of wavelength $\lambda=633$\,nm (also the case for all other experiments detailed below). The PLM chip is imaged, using a pair of lenses in a 4-f configuration, onto a mask with two pinholes placed side by side. A CMOS camera (Cam 1, Basler acA640-300gm) images the Fourier plane of the pinholes, forming interference fringes on the camera (similar to those observed in Young's double slit experiment), as shown in the inset of Fig.~\ref{fig:calib}(d).

By pistoning the micro-mirrors on one side of the PLM -- that are imaged through just one of the pinholes -- we adjust the phase of light emanating from one pinhole with respect to the other, and so observe the fringes on the camera shift laterally. We track the phase of the fringe motion as a function of micro-mirror state, which directly measures $\theta(i)$. As before, we can repeat this measurement for different mirror bias voltages to tune the PLM response to the wavelength of the illumination. We find a similar lookup table to our image plane calibration using this Fourier plane calibration method (in this case with a mirror bias of $+0.1$\,V), as shown in Fig.~\ref{fig:calib}(c). We also note that our image plane and Fourier plane lookup tables were measured with two different PLMs in two different labs -- highlighting similarity of the PLMs and the repeatability of the methods used to calibrate them.

\section{Diffraction efficiency}
A key metric quantifying SLM performance is the efficiency with which light can be steered into the first diffraction order. Once the lookup table and mirror bias are optimised, we study the diffraction efficiency of phase light modulation as a function of diffraction angle. To steer an incident collimated beam along a wavevector of $\bm{k} = [k_x, k_y, k_z]$ (representing the components of the wavevector in 3-dimensional Cartesian coordinates), we encode a hologram consisting of a linear phase ramp, the phase profile of which is given by ${\theta_{\text{tilt}}(x,y) = k_xx+k_yy}$. Here $x$ and $y$ are the Cartesian coordinates in the plane of the chip, and $z$ is normal to the chip (see Fig.~\ref{fig:mainfig}(b)). The angle of the diffracted beam with respect to the $z$-axis (also referred to as the `carrier tilt') is given by ${\phi=k_{\text{r}}/k}$ (assuming the small angle approximation), where ${k = 2\pi/\lambda = (k_x^2 + k_y^2 + k_z^2)^{1/2}}$, and $k_{\text{r}}$ is the radial k-vector: ${k_{\text{r}} = (k_x^2 + k_y^2)^{1/2}}$.

As the displayed hologram takes values between ${0 - 2\pi}$\,rad, it contains phase wrapping lines where the phase of the linear phase ramp passes through an integer multiple of $2\pi$\,rad. An important parameter when assessing the efficiency of an SLM is the number of pixels between phase wraps, $n_{\text{wrap}}$, which is given by ${n_{\text{wrap}} = 2\pi/(pk_r)}$, where $p$ is the pixel pitch (here $p=10.8$\,\SI{}{\micro\meter}). As the diffraction angle $\phi$ is increased, $n_{\text{wrap}}$ is decreased and the pixellated nature of the modulator becomes more apparent, since the ideal linear phase ramp is less well-approximated by the more coarsely stepped function displayed across the SLM. This effect leads to a reduction in the diffraction efficiency for light diffracted to larger diffraction angles.

Figure~\ref{fig:calib}(e) shows the experimentally measured diffraction efficiency as a function of diffraction angle and $n_{\text{wrap}}$. For this measurement we construct composite high dynamic range images of the diffracted orders, and remove camera dark counts at the longest exposure by subtracting one grey level from all images. We record the intensity of light transmitted into the nine diffraction orders with a camera (orders: +4, +3, +2, +1, 0, -1, -2, -3, -4) and calculate the fraction of this light transmitted into the target first diffraction order. We see that the diffraction efficiency is above $\eta = 0.8$ up to a diffraction angle of $\phi\sim8$\,\SI{}{\micro\radian} ($n_{\text{wrap}}$\,$\sim$\,6 pixels).

We note that this measurement compares the relative intensity of light transmitted into different diffraction orders, but does not account for reflection losses or light scattered from the edges of the pixels themselves --- which was recently studied~\cite{byrum2024optimizing}. Our measurements are in good agreement with other recent work studying the diffraction efficiency of a PLM~\cite{ketchum2021diffraction}, and show that a bit depth of $2^4$ yields a diffraction efficiency approaching that of 8-bit liquid crystal SLMs.

\section{In-situ aberration correction}\label{sec:ab}
We next demonstrate how a PLM can rapidly measure and correct for aberrations in an optical system. Aberrations may arise from the finite fidelity of the mode generated by the laser source, minor optical system misalignments and imperfect optical elements. In dynamic beam shaping applications, the curvature of the SLM chip itself is often a significant source of aberrations.

\begin{figure*}[t]
\centering\includegraphics[width=\linewidth]{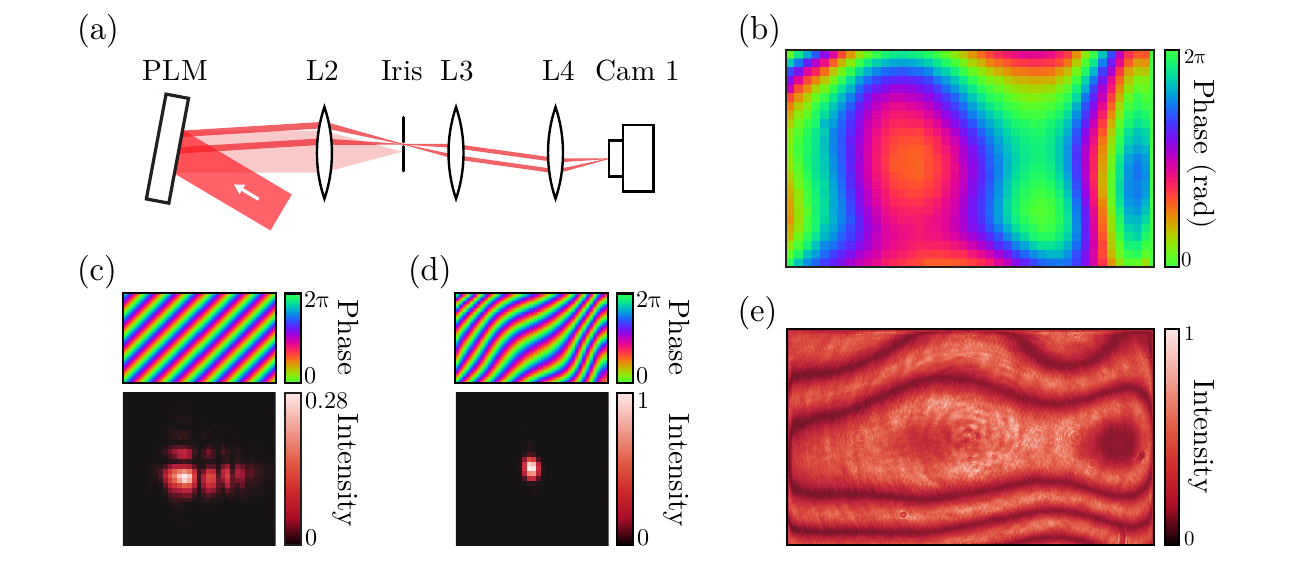}
\caption{{\bf In-situ aberration correction}. (a) Experimental setup. Beam path of a single measurement is highlighted. Lens focal lengths are the same as labelled in Fig.~\ref{fig:calib}(d). (b) A typical PLM aberration map measured with a super-pixel size of $32\times32$ micro-mirrors. (c) A distorted focus is observed in the first diffraction order (lower panel) before aberration correction is applied, when a linear phase ramp is displayed on the PLM (upper panel). (d) By encoding the phase conjugate of the measured aberration map (b) into the phase ramp (upper panel), the aberrations are removed and a diffraction limited focused spot is observed (lower panel), increasing the peak intensity by a factor of $\sim$3.5. (e) The curvature of the PLM chip is also evident from the interference pattern observed in the Twyman-Green interferometer (Fig.~\ref{fig:calib}(a)).}
\label{fig:ab}
\end{figure*}

We use in-situ wavefront correction to measure all aberrations in the beam path from the laser source up to a camera placed in the Fourier plane of the PLM~\cite{vcivzmar2010situ}. Figure~\ref{fig:ab}(a) shows the setup for this experiment. We divide the PLM into a grid of $25\times42 = 1050$ super-pixels, each of size $32\times 32$ micro-mirrors. The relative phase of light emanating from each super-pixel is sequentially measured relative to a standard reference super-pixel using phase shifting interferometry. For each measurement, light from within two super-pixels (signal and reference super-pixel) is sent into the first diffraction order and interferes on the camera placed in the Fourier plane of the PLM chip. Light reflecting from all other parts of the PLM chip is sent to the zero diffraction order where it is blocked by the iris. The signal is recorded with a single camera pixel.
Figure~\ref{fig:ab}(b) shows the resulting map of phase aberrations across the PLM, described by phase function $\theta_{\text{abb}}(x,y)$. To measure this map we cycle the reference super-pixel through 4 phase steps for each mode (0, $\pi/2$, $\pi$, $3\pi/2$), requiring the display of ${1050\times 4 = 4200}$ holograms which can be continuously streamed to the PLM in 175 frames. When operating the PLM at 1.44\,kHz, these measurements are achieved in $\sim$3\,s.

Once the aberrations are measured, they can be cancelled out by displaying the phase conjugate of the aberration map on the PLM. To create an aberration corrected focus in the first diffraction order, we display the phase function ${\theta_{\text{corr}}=\theta_{\text{tilt}}-\theta_{\text{abb}}}$. Figure~\ref{fig:ab}(c) shows the distorted spot without aberration correction, and Fig.~\ref{fig:ab}(d) shows the improvement in the spot quality when aberration correction is implemented: the light is now concentrated into a diffraction limited spot, in this case enhancing the peak intensity by a factor of $\sim$3.5. We see the spot is slightly more tightly focussed in the horizontal direction, which is to be expected since here we overfill the rectangular PLM chip with a Gaussian illumination beam, and so the beam aperture (represented by the PLM itself) is wider than it is tall.

The aberrations introduced by the PLM itself can also be directly visualised in the Twyman-Green interferometer setup (Fig.~\ref{fig:calib}(a)): when all micro-mirrors are set to the minimum height, the contours of the resulting interference pattern reveal the extent of the curvature of the PLM chip, as shown in Fig.~\ref{fig:ab}(e) -- although elevations and depressions cannot be distinguished~\cite{furhapter2005spiral}. Laterally traversing between two adjacent dark fringes in Fig.~\ref{fig:ab}(e) represents a change in the height of the PLM chip of $\lambda$. The aberrations shown in Fig.~\ref{fig:calib}(b) and Fig.~\ref{fig:calib}(e) are measured on two different PLMs. While not precisely matching, they indicate similar levels of PLM chip curvature.

\section{High-speed wavefront shaping through complex media}

Finally, we investigate the generation of complex wavefronts for the control of light transport through scattering media. When light propagates through a multiple scattering complex medium, rather than being weakly perturbed (as is the case for aberration correction in Fig.~\ref{fig:calib}), the wavefront is fragmented and the optical field is typically scrambled into a speckle pattern. This means the spatial information carried by the light is unrecognisably distorted, preventing imaging through opaque media such as frosted glass or biological tissue, even when some of the light has been transmitted~\cite{bertolotti2022imaging}. However, if the medium is static, and the scattering is purely elastic, then the light scattering process is deterministic and linear in the optical field. In this case, digital light shaping technology can be used to reverse scattering effects, by {\it pre-aberrating} an incident light field so that it evolves into a user-defined target field after transmission through the medium -- a technique known as {\it wavefront shaping}~\cite{vellekoop2007focusing}.

A versatile way to achieve this form of light control is to first create a digital model of the medium, by measuring part of its optical response function known as the {\it transmission matrix} (TM)~\cite{popoff2009measuring}. The TM is a linear matrix operator relating input fields incident on one side of a medium to the corresponding fields exiting the medium on the opposite side. It is constructed by sequentially illuminating the medium with a set of ‘probe’ light fields, and holographically measuring the resulting transmitted fields. Each measured output field is vectorised and forms one column of the TM~\cite{li2021compressively}. Once characterised, the TM $\VecT$ encodes how any linear combination of these probe fields will be transformed by the medium: $\Vecu_{\text{out}} = \VecT\Vecu_{\text{in}}$, where complex column vectors $\Vecu_{\text{in}}$ and $\Vecu_{\text{out}}$ represent an arbitrary input field, and the corresponding output field, respectively.

Knowledge of the TM enables calculation of the required shape of an incident light field to generate a specified output field on the other side -- thus enabling, for example, scanning imaging through complex media. The TM of a complex medium is typically very sensitive to changes in the medium's configuration. Therefore, the medium must remain static for at least the time taken to measure the TM and transmit the pre-corrected field. Because of this, there is a strong push for the development of fast SLMs to achieve high-speed wavefront shaping~\cite{tzang2019wavefront,feldkhun2019focusing,mounaix2020time,panuski2022full}, and PLMs open up new opportunities in this area.

\begin{figure*}[b]
\centering\includegraphics[width=\linewidth]{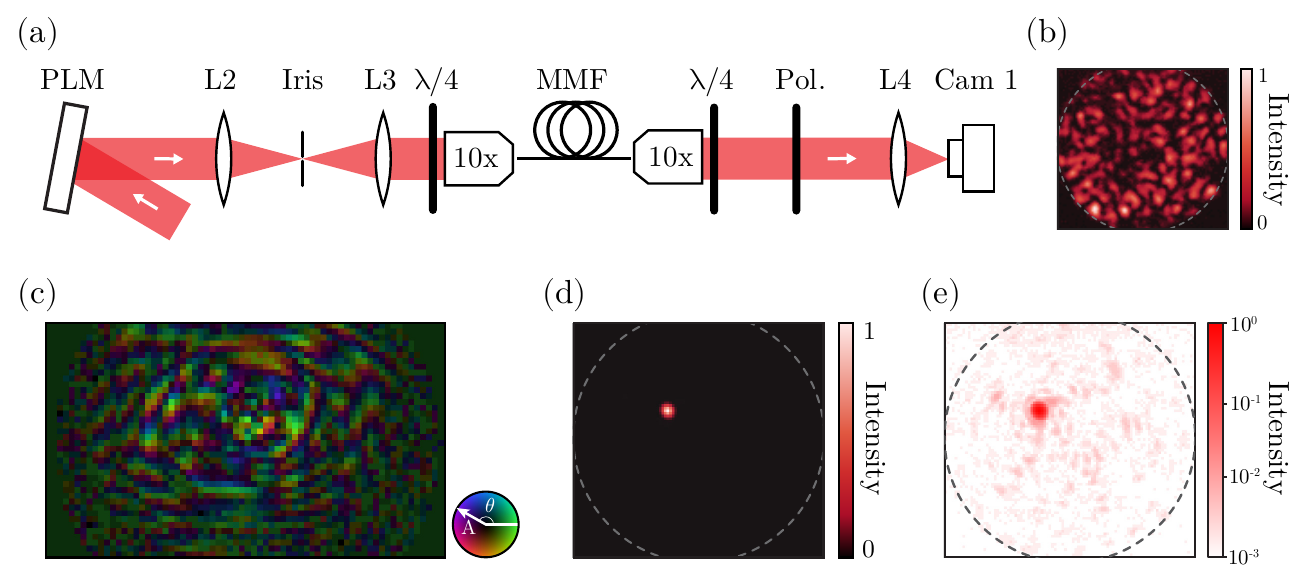}
\caption{{\bf High-speed wavefront shaping through a multimode fibre}. (a) Experimental setup. The PLM is imaged onto the pupil of an objective lens focussing light into an MMF. Light exiting the MMF is imaged onto a high-speed camera synchronised with the PLM. This setup enables measurement of the TM of the MMF, and subsequent focussing of light at its output facet. Lens focal lengths are the same as labelled in Fig.~\ref{fig:calib}(d). (b) An example of a speckle pattern at the output of the MMF when illuminating the input facet with a truncated plane-wave, showing the complete spatial scrambling of the transmitted light field. (c) The complex optical field on the plane of the PLM that will evolve into a focus at the MMF output facet, predicted from the measured TM. We encode this complex field into hologram displayed on the PLM. (d) Experimentally measured focussed point of light generated at the output of the MMF. Dotted line indicates core-cladding boundary of the MMF. (e) A high dynamic range composite image of the same focussed point.}
\label{fig:tm}
\end{figure*}

Here we test the performance of a PLM for high-speed wavefront shaping through a multimode optical fibre (MMF). Hair-thin strands of MMF support the transmission of hundreds to thousands of spatial modes, each recognisable as a unique optical field pattern of guided light. Each spatial mode is capable of acting as an independent information channel, and this ultra-high-density information capacity has motivated much interest in optical data and image transmission through MMFs~\cite{cao2023controlling}. In particular, MMFs offer an attractive platform for new types of ultra-thin micro-endoscope to image inside the body at the tip of a needle~\cite{stibuurek2023110}, and for space-division multiplexing through short-range optical interconnects in data centers~\cite{lim2023efficient}. However, spatial mode dispersion and mode coupling scramble light transmitted through MMFs, rendering them an archetypal complex medium~\cite{gigan2022roadmap}.

Figure~\ref{fig:tm}(a) shows a schematic of our optical setup for this experiment. The PLM chip is imaged onto the pupil of a $10\times$ magnification objective lens, which couples light into a 1\,m long step-index MMF with a core radius of $r=25\,$\SI{}{\micro\meter} and a numerical aperture of $\text{NA}=0.22$. This MMF supports ${N\sim(\pi r \text{NA}/\lambda)^2 \sim 750}$ spatial modes per polarisation channel. A quarter-wave plate converts the polarisation of light transmitted through the MMF to a circularly polarised state, which has been shown to suppress the polarisation coupling during transmission through short sections of step-index MMF~\cite{ploschner2015seeing} -- meaning it is not necessary to measure a polarisation resolved TM~\cite{mounaix2019control}. The output facet of the MMF is imaged onto a high-speed camera synchronised with the PLM.

To measure the TM of the MMF, we once again work in the super-pixel basis, and use an extension of the in-situ aberration correction technique applied in Sec.~\ref{sec:ab}~\cite{vcivzmar2010situ}. We sequentially illuminate the fibre with light from each super-pixel on the PLM in turn. Light emanating from a particular super-pixel results in a truncated plane-wave with a unique wavevector entering the fibre. In this case the PLM is divided into 2680 super-pixels, each of size $20\times20$ micro-mirrors -- although we only need to scan through super-pixels that fall within the disk representing the NA of the MMF on the PLM.

We employ phase-shifting holography to measure the phase of the optical field emerging from the other end of the fibre: we measure the interference between each probe beam and a reference beam that co-propagates through the fibre, once again created by emitting light from a second `reference' super-pixel also displayed on the PLM~\cite{popoff2009measuring}. The intensity profile of each trasmitted mode is measured by scanning through the probe beams a second time without the reference beam. Figure~\ref{fig:tm}(b) shows an example of a speckle pattern at the output of the fibre when it is illuminated with a single truncated plane wave. In this case TM measurement requires the display of $2680\times4=10720$ holograms, which takes $\sim$7.5\,s when operating the PLM at 1.44\,kHz. This time could be decreased to $\sim$2\,s by using off-axis digital holography instead of phase shifting holography~\cite{carpenter2022digholo}, and further reduced to below a second by employing compressive sampling techniques if some prior knowledge of the MMF under test is available~\cite{li2021compressively}.

Once the TM is measured, we use it to calculate an input light field that focuses to a target location on the output facet of the optical fibre. The required input field is given, in the super-pixel basis defined on the PLM, by $\Vecu_{\text{in}} = \VecT^{-1}\Vecu_{\text{out}}$, where $\Vecu_{\text{out}}$ is a one-hot encoded vector specifying the location of the output camera pixel where we aim to focus. Figure~\ref{fig:tm}(c) shows the structure of the optical field that should be generated by the PLM, such that it evolves into a focused spot at the output. We create this input field by adding to it the carrier tilt phase term used in TM measurement, and encoding both the spatially varying amplitude and phase profile into the displayed hologram. Amplitude modulation is achieved in a lossy manner by diffracting light away from the first diffraction order in local regions of the hologram where we wish to reduce the intensity~\cite{davis1999encoding}. 

Figure~\ref{fig:tm}(d) demonstrates the generation of a focussed point of light at the fibre output using this wavefront shaping approach. Figure~\ref{fig:tm}(e) shows a high dynamic range composite image of the same focussed spot. We quantify the fidelity of the focusing by calculating the spot {\it power ratio}, $p_{\text{r}}$, given by the ratio of the power concentrated into the diffraction limited focus to the total power transmitted through the fibre. We find the PLM is capable of high-fidelity control of light through complex media: with $p_{\text{r}}\sim0.75$, which is on-par with state-of-the-art wavefront shaping through MMFs using liquid crystal SLMs or DMDs~\cite{turtaev2017comparison,gomes2022near}. Finally, we scan 2160 points across the output facet of the MMF at 1.44\,kHz, by continuously streaming 90 unique frames of data to the PLM, each holding 24 holograms predesigned to focus to different locations at the fibre output facet. An image of each focussed point is captured with a high-speed the camera synchronised with the PLM. Supplementary movie 1 shows the scanning spot in slow-motion and in real-time.

\section{Troubleshooting}
As PLMs are a new type of SLM that is still under development, we now provide some technical details about issues that we faced during operation, and how these were resolved.

\subsection{High-speed display synchronisation}
Current PLM models have limited on-board memory, and therefore it is not possible to preload many holograms for display (as is the case with some models of DMD~\cite{stellinga2021time}). Instead, hologram data is streamed over HDMI or DisplayPort through a video interface. It is critical to synchronise the rendering of the computer's display with the PLM readout from the video screen. Without such synchronisation, hologram display is highly unreliable. Prior to implementing an appropriate continuous synchronisation scheme, we found the hologram display to be stable when commanding the PLM to play each monitor frame once (i.e.\ a burst of 24 holograms), and adding a delay of $\sim$200\,ms between monitor frames. However this approach slows down the average hologram display rate to $\sim$100\,Hz.

To overcome this issue and achieve high-speed synchronisation, we developed software that utilizes a low-level graphics API (we used Microsoft's DirectX 11) to display a full-screen texture on the PLM display monitor. Additionaly, we enable {\it vertical synchronisation} (V-Sync: a display technology used to prevent screen tearing in graphics-intensive applications) and update the RGB frames containing the bit-packed holograms during the main rendering loop, but before the frame buffer swap (which is controlled by the V-Sync signal). See the Software availability Sec.~10 for a link to our custom software package. We found that this approach enabled stable continuous hologram display at 720\,Hz using the HDMI connection on all systems tested (Exeter and Nottingham). We found that this approach also worked at 1.44\,kHz using the DisplayPort connection. However, we have observed occasional frame drops when running the system at the 1.44\,kHz in one of our systems, and are investigating the cause of this. We will update our progress resolving this issue in the documentation in our code repository.

\subsection{Data ordering}
When initialising the PLM using the LightCrafterDLPC900 graphical user interface, we used {\it Video Pattern Mode} operation mode, which enables the PLM hologram bit unpacking order to be set. It is important to pay close attention to both the bit-packing order of the holograms within each frame, and the arrangement of the bits within each memory cell defining the state of each micro-mirror. During lookup table calibration, we found that the 16 micro-mirror levels needed to be reordered so that the imparted phase monotonically increased. We found this ordering was also dependent on the value of the mirror bias voltage.

\subsection{Inactive pixels}
Reported fractions of inactive pixels in PLMs are very low~\cite{bartlett2021recent,douglass2022reliability,byrum2024optimizing}. However, in one of our PLM models we initially observed a number of columns of inactive pixels across one half of the PLM screen. We found this to be caused by small amounts of debris disrupting the ribbon connection between the control board and the PLM. This problem was fully resolved by cleaning the connectors with isopropanol.

\section{Discussion and conclusions}
We have demonstrated that PLMs can achieve continuous light-efficient and high-fidelity wavefront shaping at switching rates up to 1.44\,kHz. As PLMs are based on MEMS technology, switching rates of 10\,kHz and beyond are expected to become available in the near future.

Furthermore, PLMs have several other advantages over existing SLM technologies. For example, a key issue faced by liquid crystal SLMs is the cross-talk between adjacent pixels: the orientation of the liquid crystal in one pixel affects the liquid crystal orientation of neighbouring pixels. This lowers the fidelity of beam shaping, especially when displaying complex patterns that vary substantially from pixel to pixel~\cite{moser2019model}. In contrast, MEMS-based SLMs grant precise independent control of the state of every micro-mirror, with minimal influence on the state of surrounding mirrors (unlike membrane deformable mirrors, which suffer from actuator coupling and hysteresis). That said, PLMs may exhibit a different form of pixel coupling -- the vertical height of one micro-mirror can affect the light reflecting from neighbouring micro-mirrors due to a change in the degree to which the micro-mirror shadows those around it. Nonetheless, given that the maximum vertical micro-mirror displacement is only $\sim$3\% of the micro-mirror pitch, we expect this to be a very minor effect, that can be minimised by illuminating the PLM at close to normal incidence.

Other complications when using liquid crystal SLMs include phase flicker (unwanted oscillations in the phase response of the layer -- which are more pronounced for faster switching versions), polarisation sensitivity (only a single linear polarisation of incident light is phase modulated), long term stability of the liquid crystal layer (as it can flow over time), and the relatively low illumination power damage thresholds~\cite{carbajo2018power}. PLMs largely bypass all of these issues. Although time-dependent weak surface distortions of MEMS SLMs have been observed due to heating effects through continuous operation, these effects can be mitigated with active temperature stabilisation~\cite{rudolf2021thermal}.

Beyond structuring the phase and intensity of light beams, it should be straightforward to extend PLM operation to rapid vectorial light control~\cite{maurer2007tailoring,mitchell2016high}. Furthermore, while the spectral dispersion of MEMS-based DMDs is substantially stronger than that of liquid crystal SLMs -- a consequence of DMD micro-mirrors each tilting about their own axis~\cite{zhao2021compressive} -- the vertical displacement of PLM micro-mirrors renders their spectral dispersion similar to that of liquid crystal SLMs. We therefore expect PLMs to be compatible with relatively broadband light control~\cite{mitchell2017polarisation}. Although we note that PLMs are designed to deliver a phase delay of $2\pi$ rad to light of wavelength up to 650\,nm -- limited by the maximum range of mirror displacement~\cite{guan2022optical}, which is considerably lower than the phase range of deformable mirrors.

A drawback of PLMs is their relatively low bit depth. Moreover, as the lookup table relating the steps in micro-mirror height to the imparted phase delays is non-linear, we found that 12 micro-mirror levels are distributed over the $0-\pi$\,rad phase range, and the remaining 4 levels cover the $\pi-2\pi$\,rad range (see Fig.~\ref{fig:calib}(c)). This results in a small further reduction in the fidelity (or efficiency) of beam modulation.

PLMs have exciting potential applications to a broad range of fields. These encompass imaging and microscopy~\cite{lee2013multi,papadopoulos2017scattering,sohmen2024complex}, emerging forms of optical information processing~\cite{xia2024nonlinear,yildirim2024nonlinear,kupianskyi2023high}, and quantum photonics~\cite{lib2024high,valencia2020unscrambling}. Future developments to PLM technology that would facilitate these applications include higher levels of on-board memory allowing holograms to be pre-loaded and then played back at high speed. Real-time operation with minimal latency (currently limited to at least $\sim$17\,ms by the DisplayPort interface) is also desirable for adaptive optics and wavefront shaping through dynamic complex media~\cite{wang2015focusing}. There is also scope to improve reflection efficiency at specific wavelengths using anti-reflection coatings.

In summary, PLMs possess a unique combination of features and are poised to fill a key performance gap in available light modulation techniques. We forsee a bright future for this emerging technology.

\section{Author contributions}
DBP and GG concieved the idea for the project. JCAR led the experiments at Exeter (initial PLM testing, Fourier plane calibration, in-situ aberration correction and high-speed wavefront shaping through MMFs). TW led the Nottingham experiments (image plane calibration, high-speed PLM testing). JCAR and TW developed PLM control software. UGB implemented PLM control in LabVIEW and took the data for Fig. 1, in conjunction with JCAR. DBP, GG and JC supervised the project. DBP and JCAR wrote the paper, with editorial input from all other authors.

\section{Software availability}
Our custom PLM control software, along with its documentation and examples, is available at \href{https://github.com/structuredlightlab/plmctrl}{github.com/structuredlightlab/plmctrl}. We will continue to develop and update this software. Please feel free to contact us about it with any specific questions.

\section{Acknowledgements}
We thank Peter Savage at the University of Exeter physics workshop for rapid turnaround of a custom PLM mount. We thank EPSRC-funded summer student Jamie Graham for his work investigating the effects of PLM level quantisation. JCAR thanks the QUEX Institute for PhD funding (a collaborative enterprise between The University of Queensland and the University of Exeter). JC acknowledges financial support from the Australian Research Council (ARC) (FT220100103). GG acknowledges support from a UKRI Future Leaders Fellowship (MR/T041951/1). DP acknowledges financial support from the
European Research Council (ERC Starting grant, no.\ 804626).

\bibliography{PLM_refs}

\end{document}